\documentstyle[epsfig,12pt]{article}
\begin{document}
\title{Evolution of the EMC Effect in Light Nuclei}
\author{V.V.Burov, A.V.Molochkov and G.I.Smirnov\\
Joint Institute for Nuclear Research, Dubna, Russia}
\maketitle
\abstract{
The evolution of the EMC effect as a function of atomic mass $A$
is considered for the first time for the lightest nuclei,
$\rm D$, $^3{\rm H}$, $^3{\rm He}$ and $^4{\rm He}$, with an approach
based on the Bethe-Salpeter formalism.
We show that the pattern of the oscillation of the ratio
$r^A(x)$ = $F_2^A$/$F_2^{\rm N(D)}$ varies with $A$,
unlike the pattern for nuclei with masses $A$ $>$ 4, where only
the amplitude of the oscillation changes.\\[.2cm]
}

%
%
The understanding that nucleon structure can not be regarded
as unrelated to nuclear structure has been the main outcome of the
European Muon Collaboration (EMC)~\cite{aub83}.
The  nuclear environment modifies the nucleon partonic structure 
in such a way that the ratio of the nuclear and  deuteron structure 
functions, $r^A(x)$= $F_2^A$/$F_2^{\rm D}$, deviates from unity, 
resembling an oscillation with respect to the line  $r^A(x)$ = 1. 
Even though the $A$ dependence of the ratio  $r^A(x)$
could be explained with conventional nuclear structure 
considerations in the range  $A >$ 4~\cite{prl84},
the origin of the effect remained obscure~\cite{arne}.

The argument that the origin of the EMC effect is closely related
to saturation of the short-range binding forces in 3- or
4-nucleon systems was first presented in \cite{sm94,sm95}.
 Previous attempts to exploit this physics for explanation of the
effect failed because of both theoretical difficulties and
a lack of understanding on how the saturation 
would show up in the observables. This excuses the statement of
\cite{gomez}, that the data on $r^A(x)$ do not directly 
correlate with the binding energy per nucleon. 
The saturation, according to~\cite{sm95}, had to
manifest itself not in the amplitude of the oscillations, but in the
pattern of the $x$ dependence of $r^A(x)$, namely in the positions of
the three cross-over points $x_i$, $i$ = 1 --- 3, 
in which $r^A(x)$ = 1. Such a pattern can be clearly seen from 
new data of SLAC \cite{gomez} and NMC \cite{ama95}.
There does not 
exist any data on the EMC effect in the range of $A <$~4.
Most challenging therefore is to evaluate how the effect evolves in the
range of the lightest nuclear masses, 
$r^D(x) \to$ $r^{A=3}(x) \to$ $r^{A=4}(x)$, starting from the
basic property of nucleons to form bound states.

In the present Letter, we perform derivations of the relative
 changes in the nuclear
structure function  $F_2^A(x)$ with respect to the isoscalar nucleon one,
  $F_2^{\rm N}(x)$ = $\frac{1}{2} (F_2^{\rm p}(x)
+ F_2^{\rm n}(x))$, where $\rm p$ and $\rm n$ denote the free proton and
the free neutron respectively. On the other hand, the comparison  with 
experimental data can be  done only in terms of $r^A(x)$, obtained with 
 the deuteron structure function $F_2^{\rm D}(x)$ as a reference.
Therefore our final results will be presented for both cases.
In the considered range of $x$ ( 0.3 $< x <$ 0.9 )
the experiments (see Ref.~\cite{gomez})
are consistent with no $Q^2$ dependence of  $r^A(x)$. We perform
numerical calculations for a fixed $Q^2$ of 10 GeV$^2$.

%
%
 Our approach originates from
the Bethe-Salpeter  formalism~\cite{formalism}, which allows one
to treat nuclear binding effects by using general properties 
of nucleon Green functions~\cite{deut}. 
The approach is model independent in the sense
that it does not require any assumption about the nuclear structure
except that the nuclear fragments have a small relative energy. 
This allows us, as has been shown previously~\cite{deut}, to derive 
the deuteron structure function in the form:
\begin{eqnarray}
\!\!\!\!\!\!\! F_2^{{\rm D}}(x_{{\rm D}})=\int \frac{d^3k}{(2\pi)^3}
\left(F_2^{{\rm N}}(x_{{\rm N}})
\left(1-\frac{k_3}{m}\right)-\frac{M_{{\rm D}}-2E}{m}
x_{{\rm N}}\frac{d F_2^{{\rm N}}(x_{{\rm N}})}
{d x_{{\rm N}}}\right)\Psi^2({\bf k}),
\label{f2nr}\end{eqnarray} 
where $E$ is the on-mass-shell nucleon energy $E^2={\bf k}^2+m^2$, 
and $\bf k$ is the relative three momentum of the bound nucleons,
$M_{\rm D}$ is the mass of a deuteron.
The nucleon Bjorken variable is defined as 
$x_{{\rm N}}=x_{{\rm D}}m/(E-k_3)$.   
The function $\Psi^2({\bf k})$ is an analog of the three dimensional
momentum distribution and is directly related to the Bethe-Salpeter
(BS) vertex function for the deuteron 
$\Gamma^{{\rm D}}(P,{\bf k})=
\{\Gamma^{{\rm D}}(P,k)\}_{k_0=E-\frac{M_{{\rm D}}}{2}}$:
\begin{equation}
\Psi^2({\bf k})=\frac{m^2}{4E^2 M_{{\rm D}}(M_{{\rm D}}-2E)^2}
\overline{\Gamma}^{{\rm D}}(M_{{\rm D}},{\bf k})
\sum_s u^{s}({\bf k})\overline{u}^{s}({\bf k})\otimes
\sum_s u^{s}(-{\bf k})\overline{u}^{s}(-{\bf k})
\Gamma^{{\rm D}}(M_{{\rm D}},{\bf k})~.
\end{equation}

The two-term structure of Eq.~(\ref{f2nr}) defines a deviation of the
ratio $F_2^{\rm D}/F_2^{\rm N}$ from unity, which is generally 
considered as a signature of the modification of the nucleon structure.
As has been shown in ~\cite{deut}, the second term results from the
relative time dependence of the amplitude of  lepton deep inelastic
scattering (DIS) off a bound nucleon.
By extending the approach for light nuclei, $A = 3$,~$4$, we have 
discovered that it is the relationship between similar terms,
which is responsible for evolution of the nucleon partonic structure. 
 
 The hadronic part of the DIS amplitude (hadronic tensor) is related
 with the forward Compton scattering amplitude $T^{A}_{\mu \nu }$
by using the unitarity condition
\begin{equation}
W^{A}_{\mu \nu }(P,q)=\frac{1}{2\pi}{{\rm Im}}\, 
T^{A}_{\mu \nu }(P,q),
 \label{impart1}
\end{equation}
and $T^{A}_{\mu \nu }$ is defined as a 
product of electromagnetic currents averaged over nuclear states,
\begin{equation}
T^{A}_{\mu \nu}(P,q) =
i \int d^4x e^{ i q x }
\langle A|  {\rm T} \left(J_\mu \left(x \right)
J_\nu \left(0\right) \right)
|A \rangle.
\label{imp3}
\end{equation}
 
\noindent Starting from a field theory framework one can define the 
matrix element in terms of solutions of the $n$-nucleon Bethe-Salpeter
equation  and $n$-nucleon Green functions with 
the insertion of the $\rm T$-product of electromagnetic currents:
\begin{eqnarray}
\label{compt}T_{\mu \nu}^{A}(P,q)=&&\int d{\cal K} d{\cal K}^\prime
\overline{\Gamma}(P,{\cal K}){S_n}(P,{\cal K})
{\overline{G}_{2(n+1)}}_{\mu\nu}(q;P,{\cal K},{\cal K}^\prime)
{S_n}(P,{\cal K}^\prime)
\Gamma(P,{\cal K}^\prime),
\label{Compton}\end{eqnarray}
where ${\cal K}$ denotes a set of momenta, which  describes relative 
motion of nucleons, ${\cal K}=k_1,\dots, k_{n-1}$, 
$d{\cal K}={d^4k_n}/(2\pi)^4\dots {d^4k_{n-1}}/(2\pi)^4$, and 
$P$ is the total momentum of the nucleus.
The BS vertex function $\Gamma(P,{\cal K})$, introduced to describe
a nucleus in terms of virtual nucleon states, satisfies
the homogeneous Bethe-Salpeter equation
\begin{eqnarray}
\Gamma(P,{\cal K})=
-\int d{\cal K}^\prime
\overline{G}_{2n}(P,{\cal K},{\cal K}^\prime)
{S_n}(P,{\cal K}^\prime)
\Gamma(P,{\cal K}^\prime).
\label{BS}
\end{eqnarray}
The ${S_n}(P,{\cal K}^\prime)$ is direct product of $n$-nucleon 
propagators. 
The $\overline{G}_{2n}(P,{\cal K},{\cal K}^\prime)$ 
term denotes the irreducible truncated $n$-nucleon Green function
which is defined as follows:
\begin{equation}
\overline{G}_{2n}(P,{\cal K},{\cal K}^\prime)={S_n}^{-1}(P,{\cal K})
\delta({\cal K}-{\cal K}^\prime)-G_{2n}(P, {\cal K}, {\cal K}^\prime) ,
\end{equation}
where $G_{2n}$ is an exact $n$-nucleon Green function.
The Green function ${\overline{G}_{2(n+1)}}_{\mu\nu}$ represents Compton 
scattering of a virtual photon off a system of $n$-virtual nucleons.
All irreducible interaction corrections to the imaginary part of  
$T_{\mu \nu}^{A}$ are suppressed by powers of 
$1/(Q^2)$~\cite{deut}. This justifies consideration of the
 zeroth order term of 
${\overline{G}_{2(n+1)}}_{\mu\nu}$:
\begin{equation}
{\overline{G}_{2(n+1)}}_{\mu\nu}(q;P,{\cal K})=
\sum\limits_{i}{\overline G_4}_{\mu\nu}(q;P,k_i)
\otimes S^{-1}_{2n-1}
(k_1,\dots k_{i-1},k_{i+1}, \dots k_{n-1})\delta({\cal K}-{\cal K}^\prime)
 + O(1/Q^2),
\end{equation}
where  ${\overline G_4}_{\mu\nu}$ is connected with 
nucleon Compton amplitude: 
${\overline G_4}_{\mu\nu}(q;P,k_i)=u({\bf k_i})T^{\rm N}_{\mu\nu}(q, k_i)
\overline u({\bf k_i})$.
Then  $T_{\mu \nu}^A$ can be rewritten in terms of
the off-mass-shell $T_{\mu\nu}^{\rm N}$: 
\begin{eqnarray}
T_{\mu \nu}^A(P,q)=\int d{\cal K}
\sum\limits_{i}T^{\rm N}_{\mu\nu}(k_i,q)
\overline u({\bf }k_i)S_n(P, k_i)u({\bf k}_i) 
\overline{\Gamma}(P,{\cal K})S_n(P, {\cal K})
\Gamma(P,{\cal K}),
\label{compt0}\end{eqnarray}

Integration over ${k_i}_0$  can in principle relate $T^A_{\mu\nu}$
with on-mass-shell nucleon Compton amplitude. This can be realized
only after the singularities in nucleon propagators and the BS vertex 
functions  are taken into  account~\cite{deut}. Unlike the 
deuteron case, where singularities in the BS vertex function can 
be neglected, in this case they are connected with nucleon-nucleon 
bound states, which lie in the range of low relative momenta.
One can express the  singularities explicitly 
by introducing the ``bare'' BS vertex function $\gamma$,
 which is regular  with respect to the relative nucleon momenta:
\begin{eqnarray}
\Gamma(P,{\cal K})=-\int d{\cal K}
g_{2n}(P,{\cal K}, {\cal K}^\prime)
{S_n}(P,{\cal K}^\prime)
\gamma(P, {\cal K}^\prime)\label{bare},
\end{eqnarray} 
where $g_{2n}$ denotes the regular part of $n$-nucleon Green function at
$P^2 \rightarrow$ $M_A^2$. This function, however, contains
singularities of $m$-nucleon Green functions ($m<n$).
For example, in case of $^3{\rm He}$ the function  $g_6$ contains a pole 
of $G_4$, which corresponds to a bound deuteron and 
nucleon-nucleon continuous spectrum $g_4$:
\begin{equation}
G_4\left(\frac{2P}{3}+k,k_1,k^\prime_1\right)=
\frac{\Gamma^D(\frac{2P}{3}+k,k_1)
\overline{\Gamma}^D(\frac{2P}{3}+k,k^\prime_1)}
{\left(\frac{2P}{3}+k\right)^2
-M_D^2}
+g_4\left(\frac{2P}{3}+k,k_1,k^\prime_1\right)~.
\label{pole}\end{equation}
For ${\rm ^4He}$ one has additionally  the ${\rm ^3He}$ and 
${\rm ^3H}$ poles. 
Substituting these expressions into
Eq.~(\ref{compt0}) and integrating over the relative energy
of different nuclear fragments we derive the 
${\rm ^3He}$, ${\rm ^3H}$ and ${\rm ^4He}$  Compton amplitudes 
respectively in terms of physical amplitudes of the fragments.
Using the projection operator $g_{\mu\nu}$ and Eq.~(\ref{impart1})
one gets
$$\lim _{Q^2\rightarrow \infty }g^{\mu \nu }W^{\rm N(A)}_{\mu \nu }(P,q)
=-\frac 1x F^{\rm N(A)}_2(x)~.$$
Introducing now Bjorken variables 
$x_A=\frac{Q^2}{2P_A\cdot q}$ and
$x_{\rm N}=\frac{Q^2}{2P_{\rm N}\cdot q}$, 
we find $F^A_2$ for ${\rm ^3He}$ and 
${\rm ^3H}$ 
in the form:
\begin{eqnarray}
&&F_2^{{\rm ^3He}}(x_{{\rm ^3He}})=\label{he3}\\
&&\int\frac{d^3k}{(2\pi)^3}
\left[\frac{E_{\rm p}-{k_3}}{E_{\rm p}}F_2^{\rm p}(x_{\rm p})
+\frac{E_{\rm D}-{k_3}}{E_{\rm D}}F_2^{D}(x_D)
+\frac{\Delta^{{\rm ^3He}}_{\rm p}}{E_{\rm p}}
x_{\rm p}\frac{dF_2^{\rm p}(x_{\rm p})}
{dx_{\rm p}}+\frac{\Delta^{{\rm ^3He}}_{\rm p}}{E_{\rm D}}x_{\rm D}
\frac{dF_2^{\rm D}(x_{\rm D})}{dx_{\rm D}}\right]
{\Phi^2_{{\rm ^3He}}}({\bf k}),\nonumber
\end{eqnarray} 
\begin{eqnarray}
&&F_2^{{\rm ^3H}}(x_{{\rm ^3H}})=F_2^{{\rm ^3He}}(x_{{\rm ^3He}})|_{
p\leftrightarrow n} \label{h3}
\nonumber\end{eqnarray}  
and for ${\rm ^4He}$ in the form:
\begin{eqnarray}
&&F_2^{{\rm ^4He}}(x_{{\rm ^4He}})=\label{he4}\\
&&\int\frac{d^3k}{(2\pi)^3}
\left[\frac{E_{\rm p}-{k_3}}{E_{\rm p}}F_2^{\rm p}(x_{\rm p})
+\frac{E_{\rm ^3H}-{k_3}}{E_{\rm ^3H}}F_2^{\rm ^3H}(x_{\rm ^3H})
+\frac{\Delta^{{\rm ^4He}}_{\rm p}}{E_{\rm p}}
x_{\rm p}\frac{dF_2^{\rm p}(x_{\rm p})}
{dx_{\rm p}}+\frac{\Delta^{{\rm ^4He}}_{\rm ^p}}{E_{\rm ^3H}}x_{\rm ^3H}
\frac{dF_2^{\rm ^3H}(x_{\rm ^3H})}{dx_{\rm ^3H}}\right.\nonumber\\
&&+\left.\frac{E_{\rm n}-{k_3}}{E_{\rm n}}F_2^{\rm n}(x_{\rm n})
+\frac{E_{\rm ^3He}-{k_3}}{E_{\rm ^3He}}F_2^{\rm ^3He}(x_{\rm ^3He})
+\frac{\Delta^{{\rm ^4He}}_{\rm n}}{E_{\rm n}}
x_{\rm n}\frac{dF_2^{\rm n}(x_{\rm n})}
{dx_{\rm n}}+\frac{\Delta^{{\rm ^4He}}_{\rm ^n}}{E_{\rm ^3He}}x_{\rm ^3He}
\frac{dF_2^{\rm ^3He}(x_{\rm ^3He})}{dx_{\rm ^3He}}\right]
{\Phi^2_{{\rm ^4He}}}({\bf k}),
\nonumber\end{eqnarray}
where $\Delta^A_{\rm N}=-M_A+E_{\rm N}+E_{A-1}$ is the binding
energy of the corresponding nuclear fragment.
The three dimensional momentum distributions 
$\Phi^2_{A}({\bf k})$ are defined
via the ``bare'' Bethe-Salpeter  vertex functions. For example for 
$^3{\rm He}$ one has:
\begin{eqnarray}
&&\Phi^2_{^3{\rm He}}({\bf k})=\frac{m M_{\rm D}}{4E_{\rm p}E_{\rm D} 
M_{{\rm ^3He}}(M_{{\rm D}}-E_{\rm p}-E_{\rm D})^2}
\left\{\int \frac{d^4k_1}{(2\pi)^4}\frac{d^4k^\prime_1}{(2\pi)^4}
\overline{\gamma}^{^3{\rm He}}(P,k,k_1)
S_2\left(\frac{2P}{3}+k,k_1\right)\right. \label{phi}\\ &&\times\left.
\Gamma^D\left(\frac{2P}{3}+k,k_1\right)
\overline{\Gamma}^{\rm D}\left(\frac{2P}{3}+k,k^\prime_1\right)
S_2\left(\frac{2P}{3}+k,k^{\prime}_1\right) \otimes 
\left(\sum_s u^{s}_\alpha({\bf k})\overline{u}^{s}_\delta({\bf k})\right)
\gamma^{^3{\rm He}}(P,k,k^\prime_1)\right\}_{k_0={k_0}_{{\rm p}}}~,
\nonumber\end{eqnarray}
where ${k_0}_{{\rm p}}=\frac{M_{\rm ^3H}}{3}-E_{{\rm p}}$. 
Since presently there are no realistic solutions of the Bethe-Salpeter 
equation for a bound system of three or more nucleons, one has to use
phenomenological momentum distributions for numerical evaluations.
It is reasonable to assume that the momentum 
distributions in Eqs.~(\ref{he3}) and (\ref{he4}) can 
be related  with the distributions extracted from experimental data. 
In the calculations we make use of the 
distributions available from ~\cite{he3} and~\cite{new}. 
%
%
The contribution arising from continuous spectra 
($\rm ppn$ for $\rm ^3He$ and $\rm ppnn$ for $\rm ^4He$) is small 
in the considered kinematical range and does not change comparison
of the final result with the data. This justified some simplifications
which resulted in rather transparent form of Eqs.~(\ref{he3})
and~(\ref{he4}). The procedure has been consistently taken into
account in the normalization of the momentum distributions
 $\Phi^2_{\rm ^3He}$ and $\Phi^2_{\rm ^4He}$.

This result 
reduces to the one obtained within the 
$x$-rescaling model~\cite{xresc} and for $A$~=~3 becomes:
\begin{equation}
F_2^{\rm ^3He}(x_{\rm ^3He})=\int dy d\epsilon 
\left\{F_2^{\rm p}\left(\frac{x_{\rm ^3He}}
{y-\frac{\epsilon}{M_{\rm ^3He}}}\right)
f^{\rm p/{^3He}}(y,\epsilon) 
+ F_2^{\rm D}\left(\frac{x_{\rm ^3He}}{y-\frac{\epsilon}
{M_{\rm ^3He}}}\right)f^{\rm D/{^3He}}(y,\epsilon)\right\}~,
\label{xresc3}\end{equation} 
where $\epsilon=\Delta^{{\rm ^3He}}_{\rm p}$ has the meaning
of a nucleon (deuteron) separation energy and 
$f^{\rm p(D)/^3He}(y, \epsilon)$ are the $\rm ^3He$ spectral functions
for a bound proton (deuteron):
\begin{eqnarray}
f^{\rm p(D)/^3He}(y, \epsilon)=\int \frac{d^3k}{(2\pi)^3}
\Phi^2_{^3{\rm He}}({\bf k})
\frac{m}{E_{\rm p(D)}} y\delta\left(y-\frac{E_{\rm p(D)}-k_3}{m}\right)
\delta\left(\epsilon-(E_{\rm p}+E_{\rm D}-M_{^3He})\right).\nonumber\\
\nonumber\end{eqnarray}

%
%
Input structure functions $F_2^{\rm p(n)}(x)$ 
are introduced via parameterizations based
on the measurements of the proton and the deuteron structure functions by
DIS experiments.
We used the most recent parameterization of  $F_2^{\rm p}(x,Q^2)$
found in~\cite{smc} and fixed the value of $Q^2$ to 10 GeV$^2$.
The structure function 
$F_2^{\rm n}(x)$ is evaluated from    
$F_2^{\rm p}(x)$ and from the  ratio  $F_2^{\rm n}(x)/F_2^{\rm p}(x)$ 
determined in~\cite{bcdms}.
We have verified that the uncertainties in $F_2^{\rm p(n)}(x)$ are
suppressed in the obtained ratio $r^A(x)$ and thus can be
neglected in the considered kinematic range.

%
%
The results of the numerical calculations, which show how the free 
nucleon structure function $F_2^{\rm N}(x)$  ($A$ = 1) 
evolves to the deuteron ($A$ = 2) and
helium  ($A$ = 3 and 4) structure functions, are presented in Fig. 1(a).
The evolution, which starts from  $F_2^{\rm D}(x)$,
is shown in Fig. 1(b).  
Contrary to what is observed for nuclei with masses $A >$ 4, the pattern
of the oscillation of $r^A(x)$ changes its shape in the range of $A \leq$ 4.
The rate at which the changes occur is consistent with the
fast buildup of the short range binding forces. 

We compare our results for the ratio $F_2^{\rm ^4He}(x) / F_2^{\rm D}(x)$
with the available data from~\cite{gomez,ama95} in Fig. 2.
The position of the  cross-over point, obtained from our calculations as
$x_3$ = 0.919, is in reasonable agreement with the extrapolated data.
On the other hand, the corresponding point for $A$  = 3, $x_3$ = 0.855,
falls within the interval 0.84 -- 0.86, which is where the ratios 
$r^A(x)$ ($A >$ 4)
cross the line $r^A$ = 1~(cf. Ref.~\cite{gomez}). 
This means that the pattern of the EMC effect  observed in such dense
nuclei as metals is being reached at $A$ = 3. The larger value of $x_3$
at $A$ = 4 must be related with the anomalous binding energy of $^4$He.

%
%
A fundamental relation follows from the obtained results. Since 
binding corrections have the same form in Eqs.~(\ref{he3}) we can write
\begin{eqnarray}
I ~= ~ \int\limits_0^1 \frac{dx}{x} 
\left(F_2^{{\rm ^3He}}(x)-F_2^{{\rm ^3H}}(x)\right)~ = 
~\int\limits_0^1 \frac{dx}{x} \left(F_2^{{\rm p}}(x)
-F_2^{{\rm n}}(x)\right).
\label{A3A1}\end{eqnarray} 
The result represents the Gotfried sum $I$, which has been often
studied experimentally  from the combination of 
$F_2^{{\rm p}}(x)$ and $F_2^{{\rm D}}(x)$(cf.~Ref.~\cite{nmcGSR}). Such
a combination is equal to $I$ to within a correction  proportional
to $F_2^{\rm N}(x=0)$. Indeed, as follows from Eq.~(\ref{f2nr}), 
$$
I_D=\int\limits_0^1 \frac{dx}{x} \left(2F_2^{{\rm p}}(x)
-2F_2^{{\rm D}}(x)\right)
=~I~ - 
2\frac{\langle M_{\rm D}-2E_{\rm D}\rangle_{\rm D}}{m} 
F_2^{\rm N}(x=0)~.$$ 
Apparently, such tests cannot be performed
rigorously because $F_2^{\rm N}(x)$ is unknown at $x=0$.
On the other hand, an experiment, which used $^3$He and $^3$H targets,
would be able to measure the nucleon isospin asymmetry
independently of the model uncertainties in the binding corrections.

%
%
In conclusion, the method for the model-free calculations of the evolution 
of the nucleon structure in the lightest nuclei
has been developed as the extension of an approach
based on the Bethe-Salpeter formalism. The method allows one
to express $F_2^A(x)$ in terms of structure functions
of nuclear fragments and three dimensional momentum
distributions. As a result, $F_2^A$/$F_2^{\rm N(D)}$ 
have been evaluated numerically 
without finding solutions of Eq.~(\ref{BS}).

The obtained  pattern of distortions of the nucleon structure function 
prove that the EMC effect  in the lightest nuclei, {\rm D}, ${\rm ^3H}$,
${\rm ^3He}$ and ${\rm ^4He}$,
is basically  the manifestation of the short range binding forces 
in the nucleon parton distributions. The quantitative predictions
for ${\rm ^3He}$ and ${\rm ^4He}$ nuclei, which have to be
verified in future experiments at HERA or CEBAF, indicate that 
the  EMC effect  in heavy nuclei can  be naturally understood as
distortions of the nucleon parton distributions in $^3$He or $^3$H,
 which are modified by the nuclear density effects. 

Finally, the obtained  results prove that in the EMC effect range
(0.3 $<x<$ 0.9) the two-nucleon interactions can be considered
as the dominant mechanism
for the description of the nuclear binding forces.

We thank S.~Akulinichev, A.~Antonov, A.M.~Baldin, S.~Kulagin and
V.~Nikolaev for useful discussions. 
A.M. acknowledges the warm hospitality of the Special Research Center 
for the Subatomic Structure of Matter, Adelaide, Australia.
This work was supported in part by the RFBR grant N96-15-96423.

\begin{figure}[h]
\begin{center}
\mbox{\epsfxsize=0.54\hsize\epsffile{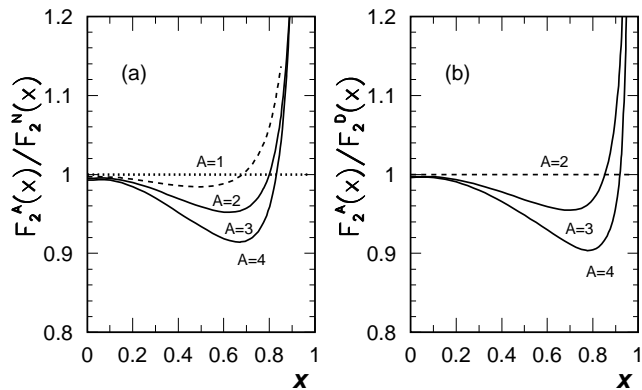}}
\end{center}

\caption{ 
(a) The ratio  $F_2^A(x)/F_2^{\rm N}(x)$.
(b)  The ratio of $F_2^A(x)$ to the deuteron
structure function $F_2^{\rm D}(x)$ ($A$ =  4) and to the combination of
structure functions  (2$F_2^{\rm D}(x)$ + $F_2^{\rm p}(x)$)/3 ($A$ =  3).
 The dashed curve in FIG. (a) shows the result of calculations,
 described in the text, for $A$ = 2. The results for $A$ = 3, 4
 are shown with the solid curves.}
\end{figure}

\begin{figure}[h]
\begin{center}
\mbox{\epsfxsize=0.4\hsize\epsffile{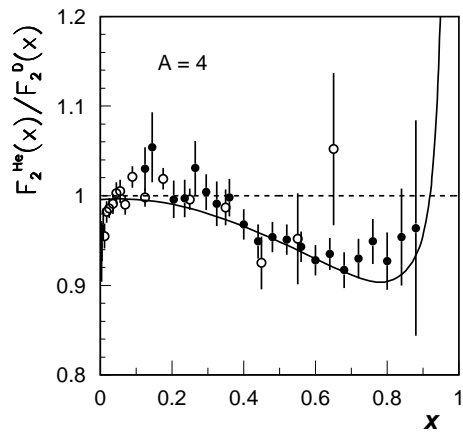}}
\end{center}

\caption{ 
The ratio $F_2^{\rm ^4He}(x)/F_2^{\rm D}(x)$.
Results of the calculation, described in the text, are shown with 
the solid curve. The data are from Ref.~[5] (filled circles) and
Ref.~[6] (empty circles).}
\end{figure}

\end{document}